\documentclass[aps,prl,twocolumn,superscriptaddress,groupedaddress,amsmath,amssymb]{revtex4-1}
\usepackage{subfigure}
\usepackage{graphicx}
\usepackage{dcolumn}
\usepackage{bm}
\usepackage{xcolor}
\usepackage{float}
\usepackage{booktabs}
\definecolor{aa}{RGB}{0,0,139}

\usepackage[colorlinks,urlcolor=aa,linkcolor=blue,anchorcolor=aa,citecolor=aa]{hyperref}

\usepackage[mathlines]{lineno}
\usepackage{multirow}
\usepackage{enumerate}
\usepackage{amsmath}
\usepackage{relsize}
\usepackage{makecell}
\def\babar{\mbox{\slshape B\kern-0.1em{\smaller A}\kern-0.1em
    B\kern-0.1em{\smaller A\kern-0.2em R}}}

\makeatletter
\renewcommand\footnote[1]{%
    \stepcounter{footnote}%
    \protected@xdef\@thefnmark{\thefootnote}%
    \@footnotemark
    \@footnotetext{#1}%
}
\makeatother

\begin{document}
\preprint{APS/123-QED}

\title{{\bf \boldmath  Observation of the $W$-annihilation process $D_s^+ \to \omega\rho^+$ and measurement of $D_s^+ \to \phi\rho^+$ in $D^+_s\to \pi^+\pi^+\pi^-\pi^0\pi^0$ decays}}

\author{
\begin{small}
\begin{center}
M.~Ablikim$^{1}$, M.~N.~Achasov$^{4,c}$, P.~Adlarson$^{76}$, O.~Afedulidis$^{3}$, X.~C.~Ai$^{81}$, R.~Aliberti$^{35}$, A.~Amoroso$^{75A,75C}$, Q.~An$^{72,58,a}$, Y.~Bai$^{57}$, O.~Bakina$^{36}$
% ...
\\
\vspace{0.2cm}
(BESIII Collaboration)\\
\vspace{0.2cm} {\it
$^{1}$ Institute of High Energy Physics, Beijing 100049, People's Republic of China\\
% ...
$^{3}$ Bochum  Ruhr-University, D-44780 Bochum, Germany\\
$^{4}$ Budker Institute of Nuclear Physics SB RAS (BINP), Novosibirsk 630090, Russia\\
% ...
$^{35}$ Johannes Gutenberg University of Mainz, Johann-Joachim-Becher-Weg 45, D-55099 Mainz, Germany\\
$^{36}$ Joint Institute for Nuclear Research, 141980 Dubna, Moscow region, Russia\\
% ...
$^{57}$ Southeast University, Nanjing 211100, People's Republic of China\\
$^{58}$ State Key Laboratory of Particle Detection and Electronics, Beijing 100049, Hefei 230026, People's Republic of China\\
% ...
$^{72}$ University of Science and Technology of China, Hefei 230026, People's Republic of China\\
% ...
$^{75}$ University of Turin and INFN, (A)University of Turin, I-10125, Turin, Italy; (B)University of Eastern Piedmont, I-15121, Alessandria, Italy; (C)INFN, I-10125, Turin, Italy\\
$^{76}$ Uppsala University, Box 516, SE-75120 Uppsala, Sweden\\
% ...
$^{81}$ Zhengzhou University, Zhengzhou 450001, People's Republic of China\\

\vspace{0.2cm}
$^{a}$ Deceased\\
% ...
$^{c}$ Also at the Novosibirsk State University, Novosibirsk, 630090, Russia\\
% ...

}

\end{center}
\vspace{0.4cm}
\end{small}
}

\vspace{0.4cm}
\begin{abstract}
    We present the first amplitude analysis and branching fraction measurement of the decay $D^+_s\to \pi^+\pi^+\pi^-\pi^0\pi^0$, using $e^+e^-$ collision data collected with the BESIII detector at center-of-mass energies between 4.128 and 4.226 GeV corresponding to an integrated luminosity of 7.33 fb$^{-1}$, and report the first observation of the pure $W$-annihilation decay $D_s^+ \to \omega\rho^+$
    with a branching fraction of $(0.99\pm0.08_{\rm stat}{\ ^{+0.05}_{-0.07}}_{\rm syst})\%$. 
    In comparison to the low significance of the $\mathcal{D}$ wave in the decay $D_s^+ \to \phi\rho^+$, the dominance of the $\mathcal{D}$ wave over the $\mathcal{S}$ and $\mathcal{P}$ waves, with a fraction of $(51.85\pm7.28_{\rm stat}{\ ^{+4.83}_{-7.90}}_{\rm syst})\%$ observed in the decay $D_s^+ \to \omega\rho^+$, provides crucial information for  the``polarization puzzle", as well as for the understanding of charm meson decays. The branching fraction of $D^+_s\to \pi^+\pi^+\pi^-\pi^0\pi^0$ is measured to be ($4.41\pm0.15_{\rm stat}\pm0.13_{\rm syst}$)\%. Moreover, the branching fraction of $D_s^+ \to \phi\rho^+$ is measured to be $(3.98\pm0.33_{\rm stat}{\ ^{+0.21}_{-0.19}}_{\rm syst})\%$, and the $R_{\phi}= {\mathcal{B}(\phi\to\pi^+\pi^-\pi^0)}/{\mathcal{B}(\phi\to K^+K^-)}$ is determined to be $(0.222\pm0.019_{\rm stat}{\ ^{+0.016}_{-0.016}}_{\rm syst}$), which is consistent with the previous measurement based on charm meson decays, but deviates from the results from $e^+e^-$ annihilation and $K$-$N$ scattering experiments by more than 3$\sigma$.
\end{abstract}

\pacs{Valid PACS appear here}
\maketitle

The polarization information of heavy-flavor mesons decaying into two vector particles ($V$) has attracted the attention of physicists for decades because of its unique advantage in the probe of new physics and novel phenomena in hadron structures~\cite{Dunietz:1990cj,Valencia:1988it}. The discrepancy between the measurement of the $B\to\phi K^{*}$ decay and the theoretical predictions, known as ``polarization puzzle", has triggered much interest in the study of $B \to VV$ decays. Various theoretical models have provided successful explanations of the phenomenon~\cite{BaBar:2003zor,Kagan:2004uw,Zou:2015iwa,Alvarez:2004ci, Yu:2024kjw}, while the situation is more debated in charm meson weak decays due to the mass of the charm quark, which is neither heavy enough to apply heavy quark symmetry, nor light enough for the application of chiral perturbation theory~\cite{Cheng:2010ry}.


In the charm sector, it is generally predicted that the transverse polarization dominates over the longitudinal one in $D_{(s)} \to VV$ decays, as indicated by the naive factorization model~\cite{ElHassanElAaoud:1999min} and the Lorentz-invariant-based symmetry model~\cite{Hiller:2013cza}. This prediction is qualitatively supported by certain experimental observations, such as $D^0 \to \bar{K}^{*0}\rho^0$~\cite{MARK-III:1991fvi}, 
but still shows quantitative discrepancies in many measurements, for example, the inability to account for the complete transverse polarization in $D^0 \to \omega\phi$~\cite{BESIII:2021raf,LHCb:2018mzv}.
A systematic approach to the polarization in $D^0 \to VV$ is proposed considering the long-distance mechanism due to the final-state interaction~\cite{Cao:2023csx}.
This approach offers a quantitatively more consistent explanation for certain polarizations observed in $D^0 \to VV$, while cases of longitudinal polarization dominance, such as in $D^0 \to \rho^0\rho^0$~\cite{FOCUS:2007ern}, still pose a puzzle. In a more detailed examination, physicists usually discuss polarizations in terms of partial-wave amplitudes with $\mathcal{S}$, $\mathcal{P}$, $\mathcal{D}$ waves corresponding to angular momentum $L=0,1,2$, respectively\footnote{The polarization in the transversity basis can be related to the $\mathcal{S}$, $\mathcal{P}$, $\mathcal{D}$ waves as Eqs.~(4.1) and (4.4) in Ref.~\cite{Cheng:2024hdo}.}. 
All models or approaches conclude that the $\mathcal{S}$ wave dominates over $\mathcal{P}$ and $\mathcal{D}$ waves. However, measurements of $D_{(s)} \to VV$ decays show that $D^0 \to K^{*-}\rho^+, \bar{K}^{*0}\rho^0, \rho^+\rho^-, \rho^0\rho^0$ are dominated by the $\mathcal{D}$ wave, and $D_s^+ \to K^{*0}\rho^+, K^{*+}\rho^0$ are dominated by the $\mathcal{P}$ wave~\cite{BESIII:2019lwn, MARK-III:1991fvi, FOCUS:2007ern,BESIII:2022bvv}. 

\begin{figure}[htp!]
\begin{center}
   \flushleft
   \begin{minipage}[t]{8.8cm}
      \includegraphics[width=0.4945\linewidth]{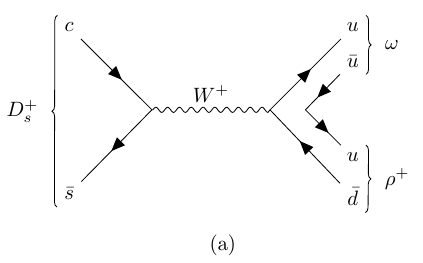}
      \includegraphics[width=0.4945\linewidth]{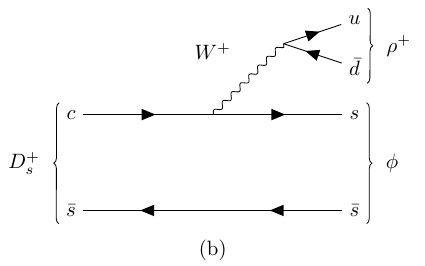}
   \end{minipage}
   \caption{Topological diagrams of (a) $D_s^+ \to \omega\rho^+$ and (b) $D_s^+ \to \phi\rho^+$. Diagram (a) also can be accessed with the replacement of $u\bar{u}$ by $d\bar{d}$.}
\label{topology}
\end{center}
\end{figure}

Polarization measurements have been comprehensively performed in $D^0$ and $D^+$ decays, but relevant measurements in $D_s^+$ decays are relatively rare. Among these, $D_s^+ \to \omega\rho^+$ stands out as one of the most important $D_s^+ \to VV$ decays to study. As a pure $W$-annihilation (WA) process, as shown in Fig.~\hyperref[topology]{1(a)}, $D_s^+ \to \omega\rho^+$ offers the best comparison with the pure external $W$-emission process $D_s^+ \to \phi\rho^+$, 
which is known to be dominated by $S$ wave. This comparison will offer an ideal approach to investigate the mechanism behind the polarization puzzle~\cite{Song:2025tog}.

Furthermore, the theoretical calculation of the WA amplitude is subject to high uncertainty due to the inaccurate estimation of the non-factorizable effects and the final-state interaction, leading to ambiguity in the predictions of the branching fractions (BFs) and \textit{CP} asymmetries of the related decays. As a result, theoretical calculations, such as the diagrammatic approach~\cite{Cheng:2024hdo,Cheng:2022vbw,Cheng:2010ry}, heavily depend on the experimental determinations of the WA amplitude as essential inputs. The small BFs of $D_s^+ \to \rho^0 \pi^+$ and $D_s^+ \to \omega\pi^+$~\cite{ParticleDataGroup:2022pth} indicate that the WA diagram is one order of magnitude smaller compared to the $W$-emission diagrams in $D \to VP$ decays, while the significantly large BFs of $D_s^+ \to a_0(980)^{+(0)}\pi^{0(+)}$~\cite{BESIII:2019jjr, BESIII:2022npc, BESIII:2021anf, BESIII:2023htx, BCKa0} and $D_s^+ \to a_0(980)^{+}\rho^{0}$~\cite{BCKa02} imply a sizeable contribution from the WA process in $D \to SP$ and $D \to SV$, where $P$ and $S$ denote pseudoscalar and scalar mesons, respectively. Up to date, no direct measurement of the WA process is available in $D \to VV$ decays.
The CLEO collaboration has measured the branching fraction of $D_s^+ \to \omega\pi^+\pi^0$ to be $(2.78\pm0.65_{\rm stat}\pm0.25_{\rm syst})\%$, and reported a relative fraction of $(0.52\pm 0.30)$ for $D_s^+ \to \omega\rho^+$ relative to the decay $D_s^+ \to \omega\pi^+\pi^0$~\cite{CLEO:2009nsf}.


In addition, the significant deviation observed in the recent BF measurements of $D_s^+ \to \phi\pi^+$ via the $\phi\to K^+K^-$~\cite{BESIII:2020ctr} and $\phi \to \pi^+\pi^-\pi^0$~\cite{BESIII:2024muy} decays indicates that the previous studies of $\phi$ decays may suffer from complexities and interferences of  backgrounds in $e^+e^-$ annihilation and $K$-$N$ scattering experiments~\cite{ParticleDataGroup:2022pth,Parrour:1975rt,Mattiuzzi:1995eze,Dolinsky:1991vq,Akhmetshin:1998se}. For the $D_s^+ \to \phi\rho^+$ decay, shown in Fig.~\hyperref[topology]{1(b)}, CLEO~\cite{CLEO:1992jqc} and BESIII~\cite{BESIII:2021qfo} have measured the BF via the channel $\phi\to K^+K^-$. The precise measurement of the decay $D_s^+ \to \phi(\to \pi^+\pi^-\pi^0)\rho^+$ together with the corresponding $\phi \to K^+K^-$ process can serve as an independent check of the BFs of the $\phi$ decays.

In this Letter, we perform the first amplitude analysis and BF measurement of $D_s^+ \to \pi^+\pi^+\pi^-\pi^0\pi^0$ using the data sets collected with the BESIII detector corresponding to a total integrated luminosity of 7.33 $\rm fb^{-1}$~\cite{BESIII:2022dxl}, and report the first observation of the pure WA decay $D_s^+ \to \omega\rho^+$ and the anomalous $\mathcal{D}$-wave dominance that deviates from the expectation of the naive factorization model~\cite{Cheng:2022vbw}. Charge-conjugate states and exchange symmetry of two identical $\pi^+$s and $\pi^0$s are implied throughout this Letter.

A description of the design and performance of the BESIII detector can be found in Ref.~\cite{BESIII:2009fln}. 
Monte Carlo (MC) events are simulated with a {\sc{geant4}}-based~\cite{GEANT4:2002zbu} detector simulation software, which includes the geometric description~\cite{Huang:2022wuo} and the response of the detector. 
Inclusive MC samples with an equivalent luminosity of 40 times that of the data are produced. They include open charm processes, initial state radiation~\cite{Kuraev:1985hb} production of vector charmonium(-like) states and the continuum processes incorporated in {\sc{kkmc}}.  The open charm processes are generated using {\sc{conexc}}~\cite{Ping:2013jka}. Final-state radiation is considered using {\sc{photos}}~\cite{Barberio:1993qi}.
In the MC generation, the known particle decays are generated with the BFs taken from the Particle Data Group (PDG)~\cite{ParticleDataGroup:2022pth} by {\sc{evtgen}}~\cite{Lange:2001uf,Ping:2008zz}, and the other modes of charmonium decays are generated using {\sc{lundcharm}} \cite{Chen:2000tv,Yang:2014vra}.

In the data samples, the $D_s^{\pm}$ mesons are produced mainly from $e^+e^- \to D_s^{*\pm}D_s^\mp \to \gamma D_s^\pm D_s^\mp$ processes. Therefore, the double-tag (DT) method~\cite{MARK-III:1985hbd, Ke:2023qzc} is employed to perform the analysis, in which a single-tag (ST) candidate is reconstructed using three hadronic decays: $D_{s}^{-}\to K_{S}^{0}K^{-}$, $D_{s}^{-}\to K^{+}K^{-}\pi^{-}$ and $D^{-}_{s}\to K^{+}K^{-}\pi^{-}\pi^{0}$, while the DT candidate is formed by selecting a $D_{s}^{+} \to \pi^+\pi^+\pi^-\pi^0\pi^0$ decay in the side of the event recoiling against the $D_s^-$ meson. The selection criteria for the final-state particles, including $K_S^0$, $K^\pm$, $\pi^\pm$, $\pi^0$, transition photon and the $D_s^-$ candidates are the same as in Ref.~\cite{BESIII:2023mie}.

For optimal resolution and to ensure that all events are within the phase space boundary, a six-constraint (6C) kinematic fit is performed. This includes the constraints of four-momentum conservation in the $e^+e^-$ center-of-mass system, as well as the constraint of the invariant mass of the tag $D_s^-$ to the known $D_s^-$ mass, and either the $D_s^+$ or $D_s^-$ candidate along with the selected transition photon to the known $D_s^{*+}$ mass, $m_{D_s^{*}}$~\cite{ParticleDataGroup:2022pth}. In cases where multiple candidates exist in an event, the one with the minimum $\chi^2$ value of the 6C kinematic fit is selected. A further kinematic fit including a seventh constraint on the mass of the signal $D_s^+$ is performed, and the updated four-momenta are used for the amplitude analysis.

\begin{figure}[htp]
  \begin{center}
    \includegraphics[width=0.4\textwidth]{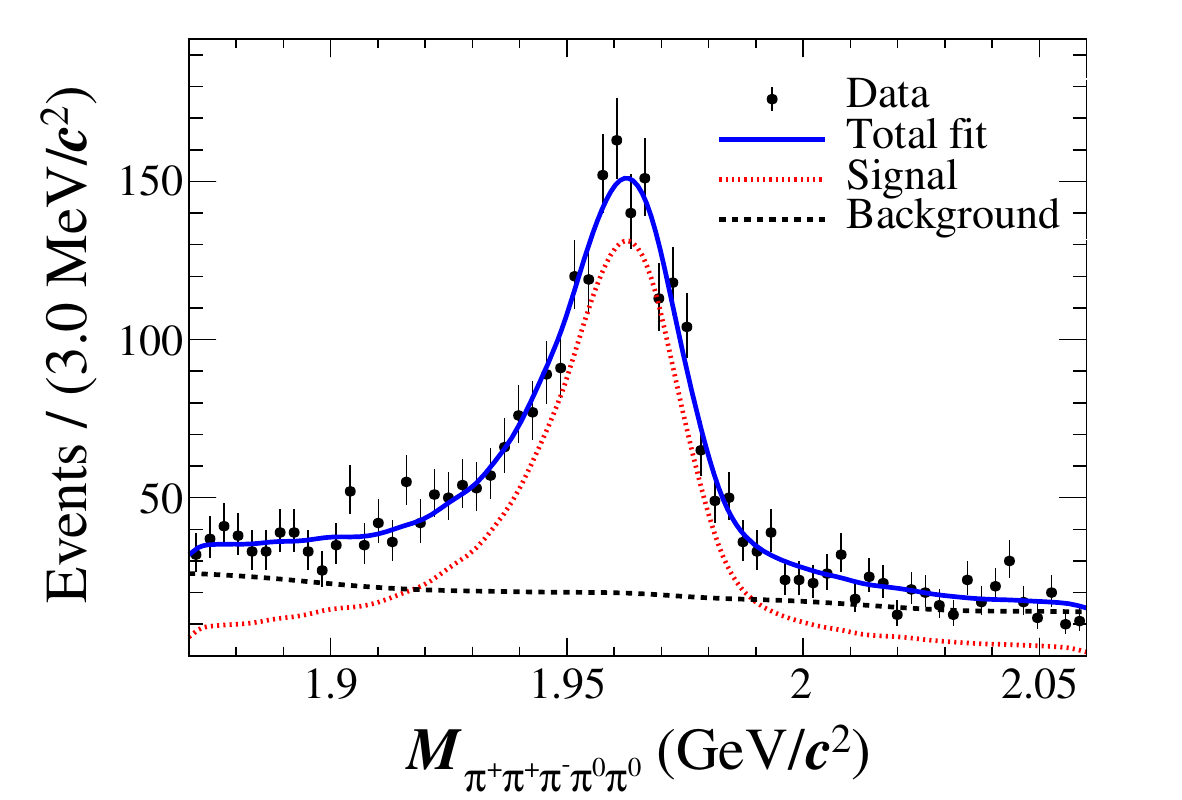}
    \caption{Fit to the invariant mass distribution of the $D_s^+$ candidates. The data are represented by points with error bars, the total fit by the blue curve,  the signal and the background components of the fit by the red dotted and the black dashed curves, respectively. 
    }
    \label{BF:fitresult}
  \end{center}
\end{figure}

To exclude the background from the $D_s^+ \to \pi^+\pi^0\eta, \eta\to\pi^+\pi^-\pi^0$ decay, the events where the invariant mass of a $\pi^+\pi^-\pi^0$ combination falls into the $\eta$ mass range $[0.49,0.58]\ {\rm GeV}/c^2$, which is about 5 times resolution, are rejected. To suppress background events from the  $K_S^0 \to \pi^0\pi^0$ decay, the invariant mass of the $\pi^0\pi^0$ combinations must be outside the 5 times resolution of $K_S^0$ corresponding to the mass range $[0.487,0.511]\ {\rm GeV}/c^2$, while, to suppress the $K_S^0 \to \pi^+\pi^-$ decays,  a secondary vertex fit~\cite{Xu:2009zzg} is performed on the $\pi^+\pi^-$ pairs, and if the ratio between the measured flight distance from the interaction point~\cite{Xu:2009zzg} and its uncertainty is larger than 2, the candidates are rejected. Another source of background comes from different open-charm processes, such as when the $D^0 \to K^-\pi^+\pi^0$ and the $\bar{D}^0\to K^+\pi^+\pi^-\pi^-$ decays are present but the first is misidentified as $D_s^- \to K^+K^-\pi^-$ and the second as $D_s^+ \to \pi^+\pi^+\pi^-\pi^0\pi^0$, in the case that the $\pi^+$ and the $\pi^0$ from the $D^0$ are wrongly exchanged with the $K^+$ and the $\pi^0$ of the $\bar{D}^0$ and an additional $\pi^0$ is selected.
This background is excluded by rejecting the events which simultaneously satisfy $|M_{K^-\pi^+\pi^0} - M_{D^0}|<30\ {\rm MeV}/c^2$ and $|M_{K^+\pi^+\pi^-\pi^-} - M_{\bar{D}^0}|<30\ {\rm MeV}/c^2$, where $M_{D^0/\bar{D}^0}$ is the known $D^0/\bar{D}^0$ mass~\cite{ParticleDataGroup:2022pth}. The analogous background from $D\bar{D}$ decays, {\it e.g.}, when $D^0\to K^-\pi^+\pi^0$ and $\bar{D}^0 \to K^+\pi^+\pi^-\pi^-\pi^0$ or $D^0\to K^-\pi^+\pi^0$ and $\bar{D}^0 \to K_S^0\pi^+\pi^-$, is excluded with the same method. To suppress the background from the $D_s^+ \to \rho^+\eta^\prime$, $\eta^\prime\to\pi^+\pi^-\gamma$ decay, we perform two kinematic fits with different decay hypotheses, assuming that the signal side $D_s^+$ decays to the signal mode or to the $\rho^+\eta^\prime,\eta^\prime\to\pi^+\pi^-\gamma$ mode; the events with the $\chi^2$ of the background hypothesis less than the $\chi^2$ for the signal one are rejected. Moreover, we require that the recoil mass $M_{\rm {rec}}$ lies in the 5 times resolution region $[1.95,2.00]\ {\rm GeV}/c^2$, defined as 
\begin{equation}
\small
    M_{\rm {rec}} = \sqrt{\left(E_{\rm cm}-\sqrt{|\vec{p}_{D_{s}^{+}\gamma}|^2c^2 + m^2_{D_s^{*}}c^4}\right)^2\!/c^4-|\vec{p}_{D_s^{+}}|^2/c^2},
\end{equation}
where $E_{\rm cm}$ is the center-of-mass energy, $\vec{p}_{D_{s}^{+}\gamma}$ is the sum of the momentum of the signal $D_s^+$ and the transition photon. We also require the energy of the transition photon in the laboratory frame to be less than 0.2 ${\rm GeV}$. Finally, we retain a sample of 1888 $D_s^+ \to \pi^+\pi^+\pi^-\pi^0\pi^0$ events with a purity of $(79.3\pm1.3)\%$ in the region [1.93, 1.99] GeV/$c^2$ of the $D_s^+$ invariant mass, determined by fitting the latter distribution of signal $D_s^+$ candidates, as shown in Fig.~\ref{BF:fitresult}. In the fit, the signal shape is the convolution of the MC signal shape and a Gaussian function, while the background shape is described with the shape obtained from the inclusive MC samples.

An unbinned maximum-likelihood method is adopted in the amplitude analysis. The probability density function is the sum of the signal amplitude and the background function with the corresponding fraction as the coefficient. The signal amplitude is parameterized with the isobar formulation in the covariant tensor formalism~\cite{Zou:2002ar}.
The total signal amplitude $\mathcal{M}$ is a coherent sum of intermediate processes $\mathcal{M} = \sum \rho_n e^{i\phi_n}\mathcal{A}_n$, where $\rho_n e^{i\phi_n}$ is the coefficient of the $n^{th}$ amplitude with magnitude $\rho_n$ and phase $\phi_n$. The $n^{th}$ amplitude $\mathcal{A}_n$ is given by the product of the Blatt-Weisskopf barrier factor of the $D_s^+$ meson $F_n^{D_s}$ and the intermediate state $F_n^i$~\cite{Blatt_Weisskopf_1973}, spin factor $S_n^i$~\cite{Zou:2002ar} and the propagator for the resonance $P_n^i$, $\mathcal{A}_n = F_n^{D_s}\prod_{i=1}^3 F_n^iS_n^iP_n^i$, where $i$ indicates the $i^{th}$ intermediate process. The relativistic Breit-Wigner (RBW) function~\cite{Jackson:1964zd} is used to describe the propagator for the resonances $\omega$, $\phi$, $a_1(1260)$ and $b_1(1235)$. The resonances $\rho$ and $\rho(1450)$ are parametrized by the Gounaris-Sakurai line shape~\cite{Gounaris:1968mw}. For $a_1(1260)$, it is considered as a quasi-three-body decay and the width is determined by integrating the amplitude squared over phase space~\cite{BESIII:2023qgj}. The masses and widths of the remaining intermediate resonances used in the fit are taken from Ref.~\cite{ParticleDataGroup:2022pth}. The background shape is estimated with inclusive MC samples using the XGBoost package~\cite{Rogozhnikov:2016bdp,Liu:2019huh}. Comparisons of events inside and outside the $D_s^+$ mass signal region for both data and MC samples indicate that the background has been well estimated. 

\begin{figure}[htp]
  \begin{center}
    \includegraphics[width=0.48\textwidth]{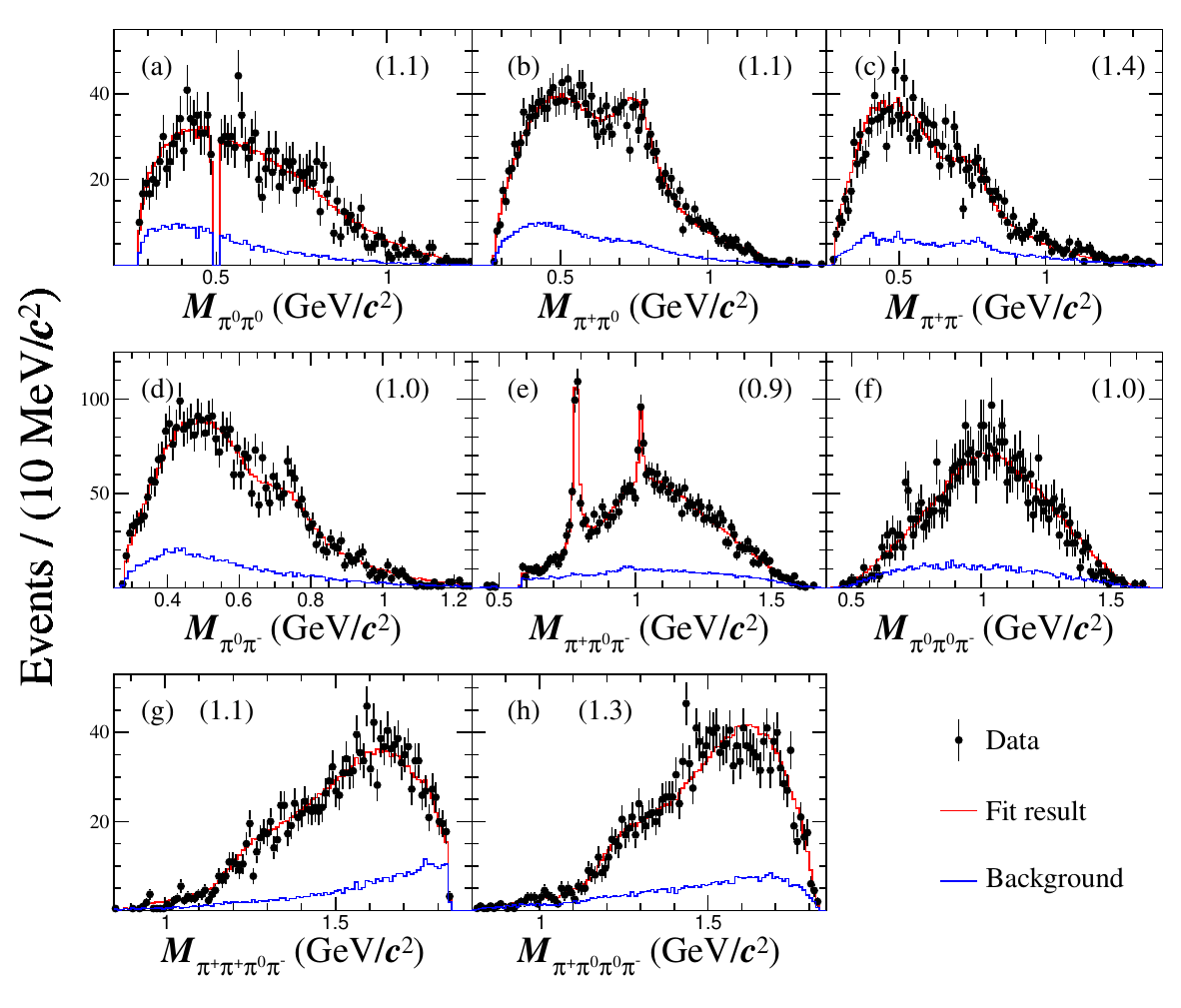}
    \caption{The projections of the fit on (a) $M_{\pi^0\pi^0}$, (b) $M_{\pi^+\pi^0}$, (c) $M_{\pi^+\pi^-}$, (d) $M_{\pi^0\pi^-}$, (e) $M_{\pi^+\pi^0\pi^-}$, (f) $M_{\pi^0\pi^0\pi^-}$, (g) $M_{\pi^+\pi^+\pi^0\pi^-}$ and (h) $M_{\pi^+\pi^0\pi^0\pi^-}$. The plots containing identical $\pi^+$ or $\pi^0$ are added into one projection. The data are represented by points with uncertainties and the fit results by the red curves. The blue curves indicate the background contribution estimated with inclusive MC samples. The numbers in brackets represent the $\chi^2/N_{\rm bin}$, where $N_{\rm bin}$ is the number of bins for each projection.
    }
    \label{pwa:fitresult}
  \end{center}
\end{figure}

The initial amplitude model is constructed with the intermediate processes which are clearly evident in the invariant mass projections, including $\omega\rho^+$ and $\phi\rho^+$. In the fit, the values of the magnitude and the phase for the dominant process $D_s^+ \to \omega\rho^+$ are fixed to be one and zero, respectively, and the other amplitudes are measured relative to this amplitude. Furthermore, the coefficients of the sub-decays of the $\phi$, $\omega$ and $a_1(1260)$ are related by Clebsch-Gordan coefficients due to the isospin symmetry. All the possible combinations with different intermediate processes are tested, and the model including the processes with statistical significance larger than $5\sigma$ is kept, where the statistical significance of each amplitude is calculated based on the change of the log-likelihood with and without this amplitude after taking the change of the degrees of freedom into account. Finally, the model with 14 amplitudes is retained. The nonresonant component is not included because its significance is less than 5$\sigma$, and including it does not improve the fit. The resolutions of narrow resonances have been considered using the same method as in Ref.~\cite{BESIII:2020ctr}. Alternative fits, leaving floating the widths of the narrow resonances, show that the obtained widths are consistent with the fixed values, indicating that the resolutions have been well assessed.
The invariant mass projections are shown in Fig.~\ref{pwa:fitresult}, while the phases, the fit fractions (FFs) and the statistical significances are listed in Table~\ref{pwa:fitresulttable}. The FF of the $n^{th}$ amplitude is calculated by 
\begin{equation}
{\rm FF}_n = \int{|\rho_n e^{i\phi_n}\mathcal{A}_n|^2}d\Phi_5/\int{|\mathcal{M}|^2}d\Phi_5,
\end{equation}
where $d\Phi_5$ is the standard element of the five-body phase space. The interference fit fractions between the amplitudes can be found in the Supplemental Material~\cite{supple}.

\begin{table*}[tp!]
	\caption{Phases, FFs, BFs and statistical significances for the amplitudes. Groups of related amplitudes are separated by horizontal lines. The last row of each group gives the total fit fraction of the above components including interference. The first and the second uncertainties in phases, FFs and BFs are statistical and systematic, respectively. The letters in bracket represent the relative orbital angular momentum between resonances. The decay chains for $\omega$ and $\phi$ are $\omega/\phi \to \pi^+\pi^-\pi^0$ (including $\rho\pi$). The BFs have been divided by the branching fractions of the decays of the final intermediate states.}
	\centering
	\setlength{\tabcolsep}{4.5pt}
	\renewcommand{\arraystretch}{1.2}
	\begin{tabular}{lc r@{ $\pm$ } c c c}
	\hline
	\hline
		Amplitude  &Phase $\phi$ (rad) &\multicolumn{2}{c}{FF (\%)} &BF (\%) &Significance\\
		\hline
		 $D_s^{+}[S]\to \omega\rho^+$  &0.0 (fixed) &6.12 &1.34  \makecell[r]{$+$0.44\\ $-$0.52} &0.30 $\pm$ 0.07 \makecell[r]{$+$0.02\\ $-$0.03} &$>10\sigma$\\
		 $D_s^{+}[P]\to \omega\rho^+$  &2.92 $\pm$ 0.13 \makecell[r]{$+$0.05\\ $-$0.07} &5.05 &0.86  \makecell[r]{$+$0.83\\ $-$0.79}  &0.25 $\pm$ 0.04 \makecell[r]{$+$0.04\\ $-$0.04}&6.1$\sigma$\\
		 $D_s^{+}[D]\to \omega\rho^+$  &4.91 $\pm$ 0.09 \makecell[r]{$+$0.04\\ $-$0.09} &10.36 &1.26 \makecell[r]{$+$0.70\\ $-$1.45} &0.52 $\pm$ 0.07 \makecell[r]{$+$0.04\\ $-$0.07} &$>10\sigma$\\
		 $D_s^{+}\to \omega\rho^+$  &- &19.98 &1.40  \makecell[r]{$+$0.92\\ $-$1.20} &0.99 $\pm$ 0.08 \makecell[r]{$+$0.05\\ $-$0.07}&-\\
		 \hline
		 $D_s^{+}[S]\to \phi\rho^+$  &0.72 $\pm$ 0.11 \makecell[r]{$+$0.06\\ $-$0.09} &11.62 &0.94  \makecell[r]{$+$0.46\\ $-$0.39} &3.32 $\pm$ 0.29 \makecell[r]{$+$0.19\\ $-$0.17}&$>10\sigma$\\
		 $D_s^{+}[P]\to \phi\rho^+$  &1.34 $\pm$ 0.15 \makecell[r]{$+$0.07\\ $-$0.30} &2.22 &0.42  \makecell[r]{$+$0.15\\ $-$0.15} &0.63 $\pm$ 0.12 \makecell[r]{$+$0.05\\ $-$0.06}&$>10\sigma$\\
		 $D_s^{+}\to \phi\rho^+$  &- & 13.86 &1.03  \makecell[r]{$+$0.47\\ $-$0.35}  &3.98 $\pm$ 0.33 \makecell[r]{$+$0.21\\ $-$0.19}&-\\
		 \hline
		 $D_s^{+}\to \rho(1450)^+\pi^0$, $\rho(1450)^+ \to \omega\pi^+$ &1.55 $\pm$ 0.11 \makecell[r]{$+$0.06\\ $-$0.08} &7.84 &0.83 \makecell[r]{$+$0.49\\ $-$0.58} &0.39 $\pm$ 0.04 \makecell[r]{$+$0.03\\ $-$0.03}&6.3$\sigma$\\
		 \hline
		 $D_s^{+}[S]\to a_1(1260)^0\rho^+$, $a_1(1260)^0[S] \to \rho^+\pi^-$  &4.61 $\pm$ 0.10 \makecell[r]{$+$0.14\\ $-$0.15} &5.19 &0.50 \makecell[r]{$+$0.22\\ $-$0.21} &0.23 $\pm$ 0.02 \makecell[r]{$+$0.01\\ $-$0.01}&$>10\sigma$\\
	     $D_s^{+}[P]\to a_1(1260)^0\rho^+$, $a_1(1260)^0[S] \to \rho^+\pi^-$  &0.06 $\pm$ 0.10 \makecell[r]{$+$0.14\\ $-$0.15} &6.25 &0.52  \makecell[r]{$+$0.23\\ $-$0.25} &0.50 $\pm$ 0.04 \makecell[r]{$+$0.02\\ $-$0.02}&$>10\sigma$\\
	     $D_s^{+} \to a_1(1260)^0\rho^+$, $a_1(1260)^0 \to \rho^+\pi^-$ &- &11.43 &0.67  \makecell[r]{$+$0.35\\ $-$0.35} &0.50 $\pm$ 0.04 \makecell[r]{$+$0.02\\ $-$0.02}&-\\
	     \hline
		 $D_s^{+}[S]\to a_1(1260)^0\rho^+$, $a_1(1260)^0[S] \to \rho^-\pi^+$  &4.61 $\pm$ 0.10 \makecell[r]{$+$0.14\\ $-$0.15} &3.64 &0.35  \makecell[r]{$+$0.17\\ $-$0.17} &0.16 $\pm$ 0.02 \makecell[r]{$+$0.01\\ $-$0.01}&$>10\sigma$\\
		 $D_s^{+}[P]\to a_1(1260)^0\rho^+$, $a_1(1260)^0[S] \to \rho^-\pi^+$ &0.06 $\pm$ 0.10 \makecell[r]{$+$0.14\\ $-$0.15} &3.76 &0.31  \makecell[r]{$+$0.20\\ $-$0.20} &0.17 $\pm$ 0.01 \makecell[r]{$+$0.01\\ $-$0.01}&$>10\sigma$\\
		 $D_s^{+} \to a_1(1260)^0\rho^+$, $a_1(1260)^0 \to \rho^-\pi^+$ &- &7.39 &0.44  \makecell[r]{$+$0.26\\ $-$0.26} &0.33 $\pm$ 0.02 \makecell[r]{$+$0.02\\ $-$0.02}&-\\
	     \hline
		 $D_s^{+}[S]\to a_1(1260)^+\rho^0$, $a_1(1260)^+[S] \to \rho^+\pi^0$  &1.85 $\pm$ 0.11 \makecell[r]{$+$0.18\\ $-$0.19} &9.43 &1.14  \makecell[r]{$+$1.13\\ $-$1.13} &0.41 $\pm$ 0.05 \makecell[r]{$+$0.05\\ $-$0.05}&9.2$\sigma$\\
		 $D_s^{+}[P]\to a_1(1260)^+\rho^0$, $a_1(1260)^+[S] \to \rho^+\pi^0$  &3.52 $\pm$ 0.12 \makecell[r]{$+$0.20\\ $-$0.21} &7.10 &0.88  \makecell[r]{$+$0.51\\ $-$0.51} &0.31 $\pm$ 0.04 \makecell[r]{$+$0.02\\ $-$0.02}&$>10\sigma$\\
		 $D_s^{+} \to a_1(1260)^+\rho^0$, $a_1(1260)^+ \to \rho^+\pi^0$ &- &16.53 &1.37  \makecell[r]{$+$1.52\\ $-$1.52} &0.73 $\pm$ 0.07 \makecell[r]{$+$0.07\\ $-$0.07}&-\\
		 \hline
		 $D_s^{+}\to b_1(1235)^+\pi^0$, $b_1(1235)^+[S] \to \omega\pi^+$  &4.27 $\pm$ 0.10 \makecell[r]{$+$0.05\\ $-$0.06} &10.79 &0.98 \makecell[r]{$+$0.68\\ $-$0.68} &0.53 $\pm$ 0.05 \makecell[r]{$+$0.03\\ $-$0.03}&9.7$\sigma$\\
		 \hline
		 $D_s^{+}\to b_1(1235)^0\pi^+$, $b_1(1235)^0[S] \to \omega\pi^0$  &1.22 $\pm$ 0.09 \makecell[r]{$+$0.04\\ $-$0.06} &14.60 &1.20  \makecell[r]{$+$0.52\\ $-$0.49} &0.72 $\pm$ 0.06 \makecell[r]{$+$0.05\\ $-$0.05}&$>10\sigma$\\
    \hline
	\hline
    \end{tabular}
		\label{pwa:fitresulttable}
\end{table*}

The systematic uncertainties for the amplitude analysis from various sources are assigned as the difference between the results from alternative fits and the nominal ones. The systematic uncertainty related to intermediate resonances is estimated by varying the uncertainties of the mass and width~\cite{ParticleDataGroup:2022pth}, and the uncertainty related to $\rho$ and $\rho(1450)$ is estimated by using the RBW function as a line shape. The barrier radii for the $D_s^+$ meson and the other intermediate states are varied by $\pm1\ ({\rm GeV}/c)^{-1}$. The uncertainty associated with the detector acceptance difference between the MC samples and data is determined by reweighting the MC events with a likelihood function according to the detector acceptance difference estimated using $e^+e^- \to K^+K^-\pi^+\pi^-(\pi^0)$ events, as in Ref.~\cite{BESIII:2020ctr}. The uncertainty related to purity differences is estimated by varying the purity within its statistical uncertainty, while for the background shape uncertainty we vary the proportion of the MC background components by $\pm30\%$. The intermediate resonances with statistical significances less than $5\sigma$ are included in the fit one by one and the largest difference with respect to the baseline fit is taken as systematic uncertainty. In addition, 100 signal MC samples are generated with the same size of data based on the amplitude model obtained in this work, and the input/output check has been done. All the fitted pull values that deviate from zero are assigned as the corresponding systematic uncertainties. The total uncertainties are determined by adding all the contributions in quadrature. The detailed results can be found in the Supplemental Material~\cite{supple}.

The BF of the $D_s^+ \to \pi^+\pi^+\pi^-\pi^0\pi^0$ decay is measured with a precise estimation of the detection efficiency based on the signal MC sample generated according to the amplitude analysis model. The BF is determined using the same tag modes and event selection criteria as in the amplitude analysis. In the measurement of the BF, a fit to the invariant mass of $D_s^\pm$ is performed in order to obtain the ST yields ($Y_{\rm tag})$ and DT yields ($Y_{\rm sig}$), together with the ST efficiencies ($\epsilon_{\rm tag}$) and DT efficiencies ($\epsilon_{\rm tag,sig}$) estimated with the corresponding signal MC samples. The BF is given by $\mathcal{B}(D_s^+ \to \pi^+\pi^+\pi^-\pi^0\pi^0) = (Y_{\rm sig}/\sum_i Y_{\rm tag}^i \epsilon_{\rm tag,sig}^i/\epsilon_{\rm tag}^i)$, where the index $i$ denotes the $i^{th}$ tag mode. The ST fit results are the same as in Ref.~\cite{BESIII:2023mie}. 
The DT fit is the same as the Fig.~\ref{BF:fitresult}. We obtain a DT yield of $1985\pm68$, thus the BF is measured to be ($4.41\pm0.15_{\rm stat}\pm0.13_{\rm syst}$)\% by dividing by the $\pi^0\to\gamma\gamma$ BF~\cite{ParticleDataGroup:2022pth}. It must be noted that the obtained BF does not include the contribution from the $D_s^+ \to \pi^+\pi^0\eta, \eta \to \pi^+\pi^-\pi^0$ decay.

For the BF measurement, the systematic uncertainty of the ST yields is estimated as in Ref.~\cite{BESIII:2023mie}. The uncertainty related to the background shape in the fit of the signal $D_s^+$ distribution is assigned by repeating the fit by changing the size of the MC background components by $\pm30\%$. The $\pi^\pm$ particle identification and tracking efficiencies and the $\pi^0$ reconstruction efficiency are studied with $e^+e^- \to K^+K^-\pi^+\pi^-(\pi^0)$ events, and the corresponding uncertainties are assigned. The systematic uncertainty from the amplitude analysis model is studied by varying the parameters in the amplitude analysis fit according to the covariance matrix. The uncertainty related to the requirements on $M_{\rm rec}$ and on the energy of the transition photon is assigned as the difference between the data and MC efficiencies in the control sample $D_s^+ \to K_S^0K^-\pi^+\pi^+$. The detailed results can be found in the Supplemental Material~\cite{supple}.

In summary, we present the first amplitude analysis and BF measurement of the decay $D_s^+ \to \pi^+\pi^+\pi^-\pi^0\pi^0$. Using the obtained FFs in Table~\ref{pwa:fitresulttable} and the measured $\mathcal{B}(D_s^+ \to \pi^+\pi^+\pi^-\pi^0\pi^0)$, the absolute BF of the intermediate states can be calculated by $\mathcal{B}_i = {\rm FF}_i \times \mathcal{B}(D_s^+ \to \pi^+\pi^+\pi^-\pi^0\pi^0)$, as listed in Table~\ref{pwa:fitresulttable}, by dividing by the BFs of the sub-decays of the intermediate resonances~\cite{ParticleDataGroup:2022pth}. The pure WA decay $D_s^+ \to \omega\rho^+$ is observed for the first time with the absolute BF to be $(0.99\pm0.08_{\rm stat}{\ ^{+0.05}_{-0.07}}_{\rm syst})\%$ and a significance larger than 10$\sigma$. The measured BF provides the first direct experimental determination on a WA process in $D\to VV$ decays. The BF of this decay is of the same order of magnitude as $D_s^+ \to a_0(980)^{+(0)}\pi^{0(+)}$ and far larger than other WA processes. In comparison to the dominance of the $\mathcal{S}$ wave and the low significance of the $\mathcal{D}$ wave in the pure external $W$-emission decay $D_s^+ \to \phi\rho^+$, the observed fraction $(51.85\pm7.28_{\rm stat}{\ ^{+4.83}_{-7.90}}_{\rm syst})\%$ for the $\mathcal{D}$ wave in $D_s^+ \to \omega \rho^{+}$ deviates from the expectation of the naive factorization model~\cite{Cheng:2022vbw}. The information on the partial-wave amplitudes of this pure WA process
can offer important insights for unraveling the ``polarization puzzle". In addition, the BF of $D_s^+\to \omega\pi^+\pi^0$ is calculated to be $(2.31\pm0.13_{\rm stat}{\ ^{+0.10}_{-0.11}}_{\rm syst})\%$ considering the interference between amplitudes, which is consistent with the CLEO measurement~\cite{CLEO:2009nsf} within 1$\sigma$.

Furthermore, the absolute BF of $D_s^+ \to \phi\rho^+$ is measured to be $(3.98\pm0.33_{\rm stat}{\ ^{+0.21}_{-0.19}}_{\rm syst})\%$ by dividing by the BF of $\phi \to \pi^+\pi^-\pi^0$~\cite{ParticleDataGroup:2022pth}. The obtained BF deviates from the value measured in $D_s^+ \to \phi(\to K^+K^-)\rho^+$~\cite{BESIII:2021qfo} by $3.1\sigma$ and from the theoretical prediction~\cite{Bedaque:1993fb} by $4.4\sigma$. Only $\mathcal{S}$ and $\mathcal{P}$ waves are observed in the nominal model. Taking the results from Ref.~\cite{BESIII:2021qfo} and this Letter, $R_{\phi}= {\mathcal{B}(\phi\to\pi^+\pi^-\pi^0)}/{\mathcal{B}(\phi\to K^+K^-)}$ is determined to be $(0.222\pm0.019_{\rm stat}{\ ^{+0.016}_{-0.016}}_{\rm syst})$, which is consistent with the value extracted from $D_s^+ \to \pi^+\pi^+\pi^-\pi^0$~\cite{BESIII:2024muy} within $1\sigma$, indicating the inconsistency between the $R_\phi$ measured in charmed hadron decays and the current PDG value. The rich structure shown in the decay $D_s^+ \to \pi^+\pi^+\pi^-\pi^0\pi^0$, along with the measured fractions of partial-wave amplitudes of $D_s^+ \to \omega \rho^{+}$ and  $D_s^+ \to \phi \rho^{+}$, provide key information for the investigation of charm meson decays and of the decays involving the $\phi$ meson.


The BESIII Collaboration thanks the staff of BEPCII and the IHEP computing center for their strong support. The authors greatly thank Professor H.~Y.~Cheng from Institute of Physics, Academia Sinica and Professor Q.~Zhao from Institute of High Energy Physics for the valuable suggestions. This work is supported in part by National Key R\&D Program of China under Contracts Nos. 2023YFA1606000, 2020YFA0406300, 2020YFA0406400; National Natural Science Foundation of China (NSFC) under Contracts Nos. 12447110, 11635010, 11735014, 11935015, 11935016, 11935018, 12025502, 12035009, 12035013, 12061131003, 12192260, 12192261, 12192262, 12192263, 12192264, 12192265, 12221005, 12225509, 12235017, 12361141819; the Chinese Academy of Sciences (CAS) Large-Scale Scientific Facility Program; the CAS Center for Excellence in Particle Physics (CCEPP); Joint Large-Scale Scientific Facility Funds of the NSFC and CAS under Contract Nos. U2032104, U1832207; The Excellent Youth Foundation of Henan Scientific Committee under Contract No.~242300421044;
100 Talents Program of CAS; The Institute of Nuclear and Particle Physics (INPAC) and Shanghai Key Laboratory for Particle Physics and Cosmology; German Research Foundation DFG under Contracts Nos. FOR5327, GRK 2149; Istituto Nazionale di Fisica Nucleare, Italy; Knut and Alice Wallenberg Foundation under Contracts Nos. 2021.0174, 2021.0299; Ministry of Development of Turkey under Contract No. DPT2006K-120470; National Research Foundation of Korea under Contract No. NRF-2022R1A2C1092335; National Science and Technology fund of Mongolia; National Science Research and Innovation Fund (NSRF) via the Program Management Unit for Human Resources \& Institutional Development, Research and Innovation of Thailand under Contracts Nos. B16F640076, B50G670107; Polish National Science Centre under Contract No. 2019/35/O/ST2/02907; Swedish Research Council under Contract No. 2019.04595; The Swedish Foundation for International Cooperation in Research and Higher Education under Contract No. CH2018-7756; U. S. Department of Energy under Contract No. DE-FG02-05ER41374.

\bibliography{references}

\end{document}